\documentstyle[12pt,worldsci]{article}
\begin{document}
\def\z2{\ifmmode Z_2\else $Z_2$\fi}
\def\ie{{\it i.e.},}
\def\eg{{\it e.g.},}
\def\etc{{\it etc}}
\def\etal{{\it et al.}}
\def\ibid{{\it ibid}.}
\def\to{\rightarrow}
\def\epem{\ifmmode e^+e^-\else $e^+e^-$\fi}
\def\Re{{\cal R \mskip-4mu \lower.1ex \hbox{\it e}\,}}
\def\Im{{\cal I \mskip-5mu \lower.1ex \hbox{\it m}\,}}
\pagestyle{empty}
\setlength{\baselineskip}{2.6ex}

\title{{\bf SIGNALS FOR TOP QUARK ANOMALOUS CHROMOMAGNETIC MOMENTS AT
COLLIDERS}}
\author{Thomas G.~Rizzo\\
%\vspace{0.3cm}
\vspace{0.3cm}
{\em Stanford Linear Accelerator Center, Stanford University, Stanford, CA
94309, USA}}
\maketitle

\begin{center}
\parbox{13.0cm}
{\begin{center} ABSTRACT \end{center}
{\small\hspace*{0.3cm}
The Tevatron and the Next Linear Collider(NLC) will be excellent tools for
probing the detailed nature of the top quark. We perform a preliminary
examination of the influence of an anomalous chromomagnetic moment for the
top, $\kappa$, on the characteristics of $t\bar t$ production at the Tevatron
and on the spectrum of gluon radiation associated with $t\bar t$ production
at the NLC. In particular, we analyze the sensitivity of future data to
non-zero values of $\kappa$ and estimate the limits that can be placed on
this parameter at the Tevatron and at the NLC with center of mass
energies of $\sqrt {s}=$ 500 and 1000 GeV. Constraints on $\kappa$ from low
energy processes, such as $b\rightarrow s\gamma$, are briefly discussed. }}

\end{center}

The probable discovery of the top quark at the Tevatron{\cite {1}} has renewed
thinking about what may be learned from a detailed study of its properties. It
is believed that the details of top quark physics may shed some light on new
physics beyond the Standard Model(SM). Amongst others, one set of the top's
properties which deserve study is its couplings to the various gauge bosons;
up until now such analyses{\cite {2}} have concentrated on the electroweak
couplings of the top. In this work we consider the possible existence of an
anomalous chromomagnetic moment, dimension-5 coupling, $\kappa$, at the
$t\bar tg$ vertex and explore the capability of the Tevatron and NLC to probe
this kind of new physics. Such interactions may arise in extended technicolor
or compositeness scenarios. At present, only rather weak limits on $\kappa$
(of order 10) exist, in particular, from operator mixing contributions to the
$b\to s\gamma$ decay. (See the last paper in Ref. 2). For details of the
analysis presented below, see Ref. 3.

At the Tevatron, both $gg,q\bar q \to t\bar t$ subprocesses are modified by
the existence of $\kappa \neq 0$ with the $q\bar q(gg)$ case displaying a
quadratic(quartic) $\kappa$ dependence. In the results below only the SM NLO
and gluon resummation corrections{\cite {4}} are incorporated by way of
`K-factors'. For $\kappa \neq 0$ the relative weights of the $gg$ and
$q\bar q$ subprocesses can be drastically altered as can be seen in Fig.1a. We
also see that the total $\sigma$ can be dramatically increased or decreased
via $\kappa \neq 0$; in particular, the CDF $\sigma$ result can be reproduced
if $\kappa \simeq 0.25$. If we assume that in the future the $t\bar t$
$\sigma$ settles down to its SM value, we can estimate the constraints that
this would impose on $\kappa$ including uncertainties(which we estimate using
Refs. 1 and 5) due to ($i$)scale
ambiguities, ($ii$)parton density variations, ($iii$)NNLO QCD corrections,
and ($iv$)the machine luminosity, as well as statistics. For ${\cal L}=
100(250,500,1000)pb^{-1}$ we obtain the $95\%$ CL ranges of $-0.14 \leq \kappa
\leq 0.15$, $-0.11 \leq \kappa \leq 0.12$, $-0.09 \leq \kappa \leq 0.11$, and
$-0.08 \leq \kappa \leq 0.11$, respectively.

One might ask if the top pair $p_t$-, rapidity($y$-), or invariant mass($M$-)
distributions can be used to increase the sensitivity to $\kappa \neq 0$; we
find that the by far dominant effect on these observables (for small values of
$\kappa$) is an approximate overall rescaling of the observable by the ratio
of the $\kappa$-dependent to the SM cross
sections. (This results from the fact that the $t\bar t$ threshold region
is found to dominate in the evaluation of $\sigma$'s.)
Almost all deviations from this simple rescaling occur at very large $M$ or
$p_t$ values where statistics will always remain quite meager for interesting
values of $\kappa$. Fig. 1b shows this situation
explicitly for the $t\bar t$ $p_t$-distribution. We conclude that the total
$t\bar t$ cross section provides the best probe of $\kappa$ at the Tevatron.

At the NLC, the $t\bar tg$ vertex can only be directly explored via the QCD
radiative process $e^+e^- \to t\bar tg${\cite {5}}. Relative to the Tevatron,
this results in a substantial loss in statistics which can be compensated for
by the cleanliness of the environment as well as a reduction in the associated
theoretical uncertainties. Since the new
$\kappa$-dependent interaction is proportional to the gluon 4-momentum, we are
thus lead to a study of the gluon energy distribution associated with
$t\bar t$. The dominant effect of $\kappa \neq 0$ is to induce an increase in
the high energy tail of this distribution. This same energy dependence leads
to the observation that the finite $\kappa$ contributions grow rapidly with
increasing $\sqrt {s}/2m_t$, implying increased sensitivity at an NLC with
$\sqrt {s}=1$ TeV instead of 500 GeV. In this first study, we ignore
effects from top decay(except in the statistics) and perform a LO analysis
following the work in Ref. 5. To reduce scale ambiguities and contributions
from higher orders, we employ the scheme of Brodsky \etal (BLM) in Ref. 6.
Estimates of contributions from these higher order are lumped into the
uncertainties when obtaining limits. Fig. 2a shows this distribution for
the case of $\sqrt {s}=1$ TeV for $\alpha_s=0.10$ while Fig. 2b shows the
result of integrating this distribution for values of
$z=2E_{glu}/{\sqrt {s}}>0.4$. Assuming that the SM results are realized, bounds
on $\kappa$ may be
obtainable by either ($i$)counting excess events with high energy gluon jets
or ($ii$)by a fit to the gluon energy distribution via a Monte Carlo analysis.
Events are selected with at least one b-tag as well as one high $p_t$ lepton
and gluon jet energies larger than 200 GeV. Such large jet energies will allow
a clean separation from the top decays and will simultaneously place us in
the region of greatest $\kappa$ sensitivity. For a luminosity of 200 $fb^{-1}$
the resulting $95\%$ CL allowed range is found to be $-1.0\leq \kappa \leq
0.25$. Substantial improvement is obtained by fitting the spectrum itself;
Fig. 2c shows the Monte Carlo generated spectrum and best fit($\kappa=0.06$)
assuming that the SM is realized. At $95\%$ CL, one now obtains the allowed
range of $-0.12\leq \kappa \leq 0.21$ for the same luminosity as above. For a
$\sqrt {s}=500$ GeV machine with an integrated luminosity of $30 fb^{-1}$,
these limits are significantly loosened; following the same procedure yields
the corresponding bounds $-1.98 \leq \kappa \leq 0.44$. A full analysis
including the effect of top decay and NLO corrections should be performed to
confirm these `first pass' results.

Both the Tevatron and the NLC provide complementary windows on the possible
anomalous chromomagnetic couplings of the top with different systematics. If
such a coupling were observed, it would provide a unique signature for new
physics beyond the Standard Model.

\vspace{1.0cm}
%
%%%%%%%%%%%%%%%%%%%%%%%%%%%%%%%%%%%%%%%%%%%%%%%%%%%%%%%
\def\MPL #1 #2 #3 {Mod.~Phys.~Lett.~{\bf#1},\ #2 (#3)}
\def\NPB #1 #2 #3 {Nucl.~Phys.~{\bf#1},\ #2 (#3)}
\def\PLB #1 #2 #3 {Phys.~Lett.~{\bf#1},\ #2 (#3)}
\def\PR #1 #2 #3 {Phys.~Rep.~{\bf#1},\ #2 (#3)}
\def\PRD #1 #2 #3 {Phys.~Rev.~{\bf#1},\ #2 (#3)}
\def\PRL #1 #2 #3 {Phys.~Rev.~Lett.~{\bf#1},\ #2 (#3)}
\def\RMP #1 #2 #3 {Rev.~Mod.~Phys.~{\bf#1},\ #2 (#3)}
\def\ZP #1 #2 #3 {Z.~Phys.~{\bf#1},\ #2 (#3)}
\def\IJMP #1 #2 #3 {Int.~J.~Mod.~Phys.~{\bf#1},\ #2 (#3)}
\bibliographystyle{unsrt}

%$   figure  captions
{%\small
%\vspace*{2.00in}
\noindent
Fig.~1: (a)NLO cross sections for the $q\bar q \to t\bar t$(dash-dotted)
and $gg \to t\bar t$(dotted) subprocesses as well as the total cross
section(solid) at the Tevatron as functions of $\kappa$ for $m_t=170$ GeV
using the CTEQ parton distribution functions. The horizontal dashed lines
provide the $\pm 1\sigma$ CDF cross section determination while the horizontal
dotted line is the D0 $95\%$ CL upper limit.
(b)$p_t$ distribution for top quark pairs produced at the Tevatron
assuming $m_t=170$ GeV and CTEQ parton densities. The solid curve is the SM
prediction and the upper(lower) dash-dotted, dashed, and dotted curves
correspond to $\kappa=1,\  0.5,\  0.25 (-1,$ $ -0.5,\  -0.25)$, respectively.

\medskip

%\vspace*{8.15in}
\noindent
Fig.~2: (a)Gluon jet energy spectrum assuming $\alpha_s=0.10$ for $m_t=175$ GeV
at a $\sqrt {s}=1$ TeV NLC. The upper(lower) dotted, dashed, and dot-dashed
curves correspond to $\kappa$ values of 3(-3), 2(-2), and 1(-1) respectively
while the solid curve is conventional QCD with $\kappa=0$.
(b)Integrated gluon energy spectrum for the same input parameters and labelings
as in Fig. 3 as a function of $\kappa$ assuming $z_{cut}=0.4$.
(c)Best fit gluon spectrum through the points generated by the Monte
Carlo analysis corresponding to $\kappa=0.06$.

 }

\end{document}